\def\ion#1#2{#1$\;${\small\rm\@Roman{#2}}\relax}
\documentclass[rnote]{aa}

\usepackage{graphicx}
\usepackage{txfonts}

\def\go{
\mathrel{\raise.3ex\hbox{$>$}\mkern-14mu\lower0.6ex\hbox{$\sim$}}
}
\def\lo{
\mathrel{\raise.3ex\hbox{$<$}\mkern-14mu\lower0.6ex\hbox{$\sim$}}
}
\begin{document}
   \title{{\em Swift} observations of the March 2011 outburst of the
   cataclysmic variable NSV 1436: a probable dwarf nova}
   \author{J.P. Osborne\inst{1}, K.L. Page\inst{1}, A.A. Henden\inst{2},
   J.-U. Ness\inst{3}, M.F. Bode\inst{4}, G.J. Schwarz\inst{5,6},
   S. Starrfield\inst{7}, J.J. Drake\inst{8}, E. Kuulkers\inst{9} and  A.P. Beardmore\inst{1} 
%         \fnmsep
          }

   \offprints{ }

   \institute{Dept. of Physics and Astronomy, University of Leicester, Leicester, LE1 7RH, UK
\and American Association of Variable Star Observers, 49 Bay State Rd., Cambridge, MA 02138-1203, USA
\and XMM-Newton Science Operations Centre, ESAC, Apartado 78, 28691 Villanueva
de la Ca{\~ n}ada, Madrid, Spain
\and Astrophysics Research Institute, Liverpool John Moores University, Liverpool CH41 1LDS
\and Department of Geology and Astronomy, West Chester University, West
Chester, PA 19383, USA
\and American Astronomical Society
\and School of Earth and Space Exploration, Arizona State University, Tempe,
AZ 85287-1404, USA
\and Harvard-Smithsonian Center for Astrophysics, 60 Garden Street, Cambridge, MA 02138, USA
\and INTEGRAL Science Operations Centre, ESAC, Apartado 78, 28691 Villanueva
de la Ca{\~ n}ada, Madrid, Spain
}              

   \date{Received ; accepted }

  \abstract{}{The March 2011 outburst of the poorly-studied cataclysmic variable NSV 1436 offered an opportunity to decide
between dwarf nova and recurrent nova classifications.}
	    {We use seven daily observations in the X-ray and UV by the Swift satellite, together with AAVSO V photometry, to
characterise the outburst and decline behaviour.}
	      {The short optical outburst coincided with a faint and relatively soft X-ray state, whereas in decline to
fainter optical magnitudes the X-ray source was harder and brighter. These attributes, and the modest optical outburst
amplitude, indicate that this was a dwarf nova outburst and not a recurrent nova. The rapid optical fading suggests an
orbital period below 2 hours.}
	      {}
   

   \titlerunning{NSV 1436}
   \authorrunning{J.P. Osborne} 
  \maketitle
%
%________________________________________________________________

\section{Introduction}

\label{intro}

In classical novae a white dwarf accretes from a binary companion until the pressure at the
base of the accumulated layer is sufficient to cause a thermonuclear runaway in the partially degenerate gas (for a
review see Bode \& Evans 2008). Recurrent novae occur for the same reason, but repeat on few-decade timescales; they have a special importance as they have been
proposed as progenitors of type 1a supernovae (e.g. Livio 2000), which are used to determine the fundamental parameters of the Universe. Rather few
recurrent novae are known; Schaefer (2010) lists just 10 Galactic examples. 

An announcement of an outburst of \object{NSV 1436} noted that it might be a recurrent nova, although also allowing that it might be a dwarf nova perhaps of the rare WZ Sge type. Here we describe a brief observing
programme with the {\em Swift} satellite (Gehrels et al. 2004) which aimed to constrain the nature of this object.

\section{Observations and Analysis}

Following AAVSO Alert Notice 434, announcing an outburst of the little-known
cataclysmic variable NSV 1436 at RA(2000) = 04 02 39.02, Dec(2000) = +42 50
46.0, {\em Swift} observed this object for approximately 2000 seconds daily
between 30 March and 5 April 2011. These data were processed using version 3.7
of the {\em Swift} software, corresponding to HEASOFT release 6.10.

As reported by the AAVSO, NSV 1436 was first seen at increased brightness on
28.9 March 2011 at magnitude 13.5 and reached a peak at 12.8 on 30.3 March. Quiescent magnitudes range from 16--19, although it was not been seen below magnitude 17 in the two months prior to this outburst. The last substantial recorded outburst was in 1948, raising the possibility that NSV~1436 is a recurrent nova or a rare WZ Sge type dwarf nova. On the other hand, it was seen at magnitude 14.5 on 9-10 March, suggesting a more common dwarf nova type.

Coincident with NSV 1436 in the {\em ROSAT} All Sky Bright Source
Catalogue (Voges et al. 1999) is \object{1RXS J040239.4+425037}, which had a
PSPC count rate of 0.14 count~s$^{-1}$. NSV 1436 is also spatially coincident with GALEX J040239.0+425045\footnote{http://aladin.u-strasbg.fr/aladin.gml}, which had near (2271$\AA$) and far (1528$\AA$) UV magnitudes of 18.40 and 18.96 respectively when observed between 25 December 2005 and 21 December 2006\footnote{http://galex.stsci.edu/GalexView/}.

During the {\em Swift} observations, the brightness of NSV 1436 declined from a visual/V magnitude of 13.0 to 15.4--16.2 by the third day, and was at the same level the following day; it was at magnitude 16.0 eight days later (values taken from the AAVSO website\footnote{http://www.aavso.org/data/lcg}).

At 1928$\AA$ ({\em Swift}-UVOT uvw2 filter; Roming et al. 2005) NSV 1436 declined from magnitude 11.8 (corrected for coincidence loss) to 15.6 during the observations, essentially following the AAVSO light curve, and having the same brightness on the last two observations.

\begin{figure}
\centering

\includegraphics[scale=0.5]{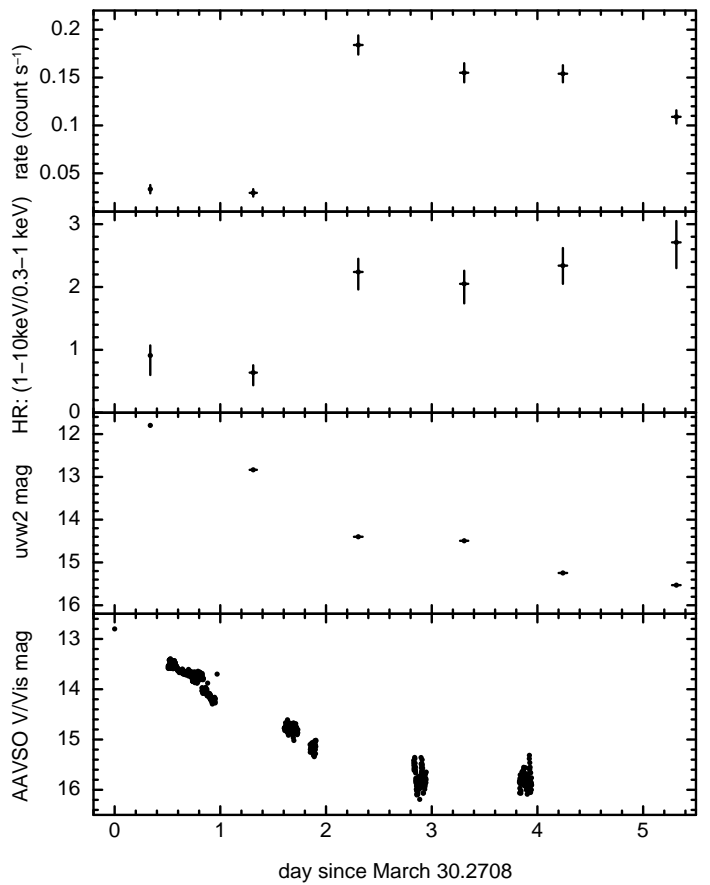}
\caption{Light curves of the March 2011 outburst of NSV 1436. From top to
bottom: {\em Swift}-XRT count rate; {\em Swift}-XRT spectral
hardness ratio; {\em Swift}-UVOT uvw2 magnitude; AAVSO visual/V magnitude.}
\label{lc}
\end{figure}

{\em Swift}-XRT (Burrows et al. 2005) detected an X-ray source at RA(2000) = 04 02 38.8, Dec(2000) = +42
50 46, 90\% error radius = 3.8$\arcsec$, consistent with the AAVSO position of
NSV 1436. The X-ray source was initially faint, at $\sim$~0.03
count~s$^{-1}$, but rose on the third day to 0.18 count~s$^{-1}$ before
declining steadily to 0.11 count~s$^{-1}$ at the end of the observations. The
(1.0-10 keV)/(0.3-1.0 keV) count rate hardness ratio was $\lo$~1 initially,
but rose to be $\go$~2 from day two onwards. 
Figure~\ref{lc} shows the light curves at X-ray, UV and optical wavelengths,
as well as the X-ray hardness ratio.

%The {\em Swift} and AAVSO data are shown in Figure~\ref{lc}.

\begin{figure}
\centering

\includegraphics[angle=-90, scale=0.36]{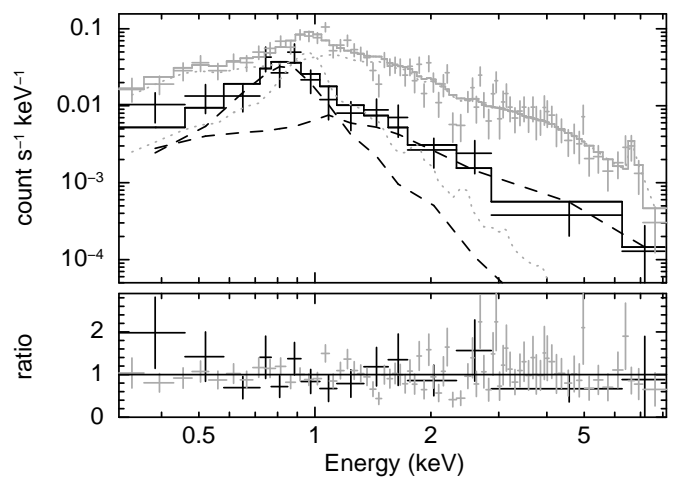}
\caption{{\em Swift}-XRT 0.3-8 keV spectral fits to the outburst (observations 1 \& 2) and quiescent (observations 3--6) spectra of NSV
1436. Both spectra require two optically thin thermal components, shown dashed
      in outburst and dotted in quiescence. NSV 1436 is brighter and hotter in quiescence than in outburst.}
\label{spec}
\end{figure}

X-ray spectral fits to data between 0.3--8 keV require two optically thin
thermal components in both the optically bright/X-ray faint and in the
optically fainter/X-ray brighter states. We fit two mekal models absorbed by a
common column which was tied between the two states: N$_{\rm
H}$=(6.5$^{+3.1}_{-2.4}$)~$\times$~10$^{20}$~cm$^{-2}$. The joint fit has a
C-statistic (Cash 1979) value of 426 for 436 degrees of freedom and is shown in Figure~\ref{spec}. Fits with single temperature models are rejected as they have C-statistic values higher by at least 40 in each state. Emission parameters are reported in Table~\ref{fit}. 

\begin{table*}
\caption{Spectral parameters}              % title of Table
\label{fit}      % is used to refer this table in the text
\centering                                      % used for centering table
\begin{tabular}{c c c c c}          % centered columns (4 columns)
\hline\hline                        % inserts double horizontal lines
NSV 1436 (Swift-XRT) &  \multicolumn{2}{c} {Optically bright/X-ray faint} & \multicolumn{2}{c}{Optically faint/X-ray}
bright \\    % table heading
& cool & hot & cool & hot\\
\hline                                   % inserts single horizontal line
kT (keV) & 0.60$^{+0.13}_{-0.15}$ & $>$2.63 & 	0.94$^{+0.15}_{-0.11}$ &
7.0$^{+3.0}_{-1.9}$ \\
Unabs. flux (10$^{-13}$ erg~cm$^{-2}$~s$^{-1}$, 0.3-10 keV) & 4.7 & 6.1 & 7.6 &	58.4\\
\hline                                             %inserts single line
\end{tabular}
\end{table*}

%Figure~\ref{lc} shows the light curves at X-ray, UV and optical wavelengths,
%as well as the X-ray hardness ratio.

The {\em ROSAT}-PSPC count rate predicted from the {\em Swift}-XRT optically
faint/X-ray bright state flux is 0.20~count~s$^{-1}$, roughly consistent with
that observed. This suggests that the {\em ROSAT} observation took place
during optical quiescence, as indeed is most likely given the apparently low outburst duty cycle.

\section{Discussion}
\label{disc}

Without optical spectroscopy it is not possible definitively to classify
NSV 1436.  The multi-temperature nature of the X-ray spectrum is a property of both novae and dwarf novae, while the lack of detection of a super-soft source with a blackbody temperature a few tens of eV (expected from unabsorbed novae, see Ness et al. 2007; Schwarz et al. 2011) is also not decisive due to the short {\em Swift} observing campaign. However, the anti-correlation of X-ray and optical flux, and the
anti-correlation of X-ray spectral hardness with optical brightness, are both characteristic of
dwarf nova outbursts (e.g. see Baskill, Wheatley \& Osborne 2005; Collins \& Wheatley 2010). 
In contrast, X-ray emission from a nova due to either shocked circumstellar gas or internal shocks within the ejecta would be expected to show cooling whilst fading after a few days, and a strong decrease in the high absorption early on (see e.g. Bode et al. 2006; Page et al. 2010); these behaviours were not seen. Finally, the outburst
amplitude of about three magnitudes is much more typical of a dwarf nova than of a nova explosion. 

The optical fading rate of dwarf nova outbursts has been shown to be almost
invariant for a given  . Using the well-defined decline rate versus
orbital period relationship of Warner (1995, fig 3.11), the rapid optical
decline of 2.5 magnitudes in two days of NSV 1436 suggests a short orbital period of P$_{\rm orb}$~$<$~2~hours. 
Although this would allow a WZ Sge-type dwarf nova classification for
NSV 1436, the modest amplitude and duration of the outburst compared to the WZ Sge types (Bailey 1979; Howell, Szkody \& Cannizzo 1995), together with the lack of reports of superhumps, which are easily detected in these systems, suggests that this   is a more common U Gem-type dwarf nova. 

\section{Summary}

Seven daily observations of the cataclysmic variable NSV 1436 by {\em Swift} during a recent outburst show that it was a faint and soft X-ray source during outburst and was harder and brighter during quiescence. This behaviour points strongly to a dwarf nova rather than a nova classification for this  . 

\listofobjects

\begin{acknowledgements}
We thank the {\em Swift} operations team and PI for their support of these observations. JPO, KLP and APB acknowledge financial support from the UK Space Agency. SS is grateful to the NSF and to NASA for partial support. 
This work made use of data supplied by the
UK Swift Science Data Centre at the University of Leicester.

\end{acknowledgements}

\end{document}